\documentclass[14pt]{article}
\usepackage[koi8-r]{inputenc}
\usepackage[saorus]{sao1}
\usepackage{graphicx}

\begin{document}
\begin{center}
{\Large \bf Stellar population of the interacting galaxies M~51.}\\
\bigskip

{\large  E.N. Tikhonov, N.A. Tikhonov$^1$, O.A. Galazutdinova$^1$$^*$.}

\bigskip

{\em $^1$ Special Astrophysical Observatory Russian Academy of Sciences,
N.Arkhyz, 369167, Russia} \end{center}

\begin{abstract}
Stellar photometry of about one million stars in the system of the interacting
galaxies M~51 (NGC~5194/NGC~5195) has been carried out on the basis of the
ACS/WFC images of the Hubble Space telescope. The distance to the system has
been determined by TRGB method and metallicity of red supergiants along the
NGC~5194 radius has been measured. A comparison of the spacial distribution
of stars in the main galaxy NGC~5194 with our empirical model of spiral
galaxies showed their similarity, in spite of interaction in system M~51.
It has been discovered that the "feathers", arising as a result of the
interaction of galaxies, mainly consist of stars of intermediate age.
\end{abstract}

\begin{center}
{\large \bf Introduction}
\end{center}

\bigskip
\noindent
The spiral galaxy M~51 (NGC~5194/5195 ore VV~001) discovered in 1773 by
Charles Messier is of great interest because it is a nearby system  of two
interacting galaxies (Fig. 1). Therefore Vorontsov-Vel'yaminov et al. (1959)
put it first in their list of interacting galaxies. M~51 system is not
far from the Local Group and it is convenient for checking of theoretical
simulation of  galaxy interaction. At present there exist two models
of galaxy interaction which describe morphology and dynamics of M~51:

   1) A model with a single parabolic approach (Toomre \& Toomre, 1972)

   2) A model of the related system with repeated passing of the companion (Salo \& Laurikainen, 2000).

In the model of Toomre and Toomre (1972) the galaxy-companion NGC~5195 crossed
the south area of the main galaxy disk NGC~5194 at a distance of $25 - 30$
kpc approximately $300 - 500$ million years ago. NGC~5195 is located now at
a distance of 50 kpc behind M~51 disk and it moves away. This model is
confirmed by the observations of kinematics of the planetary nebulae (Durrell
et al., 2003).

In the model with the repeated approaches NGC~5195 crossed first NGC~5194 disk
towards an observer, then it crossed NGC~5194 once more in the north about 100
million years ago. Now NGC~5195 is situated within a distance of about 20 kpc
behind the main galaxy NGC~5194. This model is confirmed by radio observations
of the detailed kinematics of HI (Salo \& Laurikainen, 2000).

  \vspace{0.1cm}
\begin{center}
{\large \bf Investigation of M~51 stellar population }
\end{center}
Images received at the Hubble Space Telescope with ACS/WFC in 2005 (ID 10452)
give a possibility to study the population of the galaxies directly. We
investigated 6 fields  which cover both galaxies of system M~51
(Fig. 2). Exposure durations are 1360 seconds in F555W and F814W filters.
The results of stellar photometry for almost a million stars obtained with
DAOPHOT are represented as colour --- magnitude diagram (CMD) in Fig.3a.
The obtained CMD is similar to the diagrams of normal spiral galaxies.
It shows a wide branch of blue supergiants, a branch of red supergiants and
many stars of intermediate age and red giants unresolved on the diagram.
Determinations of the distance to M~51 have been carried out many times.
The distance values obtained by different techniques are presented in the Table~1.
 \begin{table}[h]
\begin{center}
\footnotesize
\renewcommand{\tabcolsep}{4pt}
\caption{Distance to M~51}
\vspace{0.2cm}
 \begin{tabular}{|clcll|} \hline
\multicolumn{1}{|c}{Year}&
\multicolumn{1}{c}{Authors}&
\multicolumn{1}{c}{Method}&
\multicolumn{1}{c}{(m-M)}&
\multicolumn{1}{c|}{$\sigma$}\\ \hline
1974&Sandage \& Tammann&  H II regions                   &29.91 &\\
1990&Georgiev et al.              &Stellar association sizes &29.2  & 0.2\\
1994&    Iwamoto et al.           &    SN 1994Ic+O stars      &29.2  & 0.2\\
1996&      Baron et al.             &      SN 1994                    &28.9  & 0.6\\
1996&      Feldmeier               &      PNLF                        &29.62 & 0.15\\
2006&      Takats \& Vitko    &      SN 2005cs                &29.25 & 0.34\\
2007&      This paper              &     TRGB                        & 29.98 & 0.15\\
\hline
\end{tabular}
\end{center}
\end{table}

Using the photometry results we found the distance to M~51 by determining the
location of the tip of the red giant branch (TRGB method, Lee et al., 1993).
In order to decrease the influence of brighter AGB stars on the accuracy of TRGB
determination, we used three distant periphery galaxy regions, where red
giants predominate (Fig.~4). Fig.~5 shows the luminosity function of  the
red giants and the jump in luminosity function computed with the Sobel
filter determines the tip of the giant branch. Using equations of Lee et al.
(1993) we have obtained: [Fe/H] =$-$0.65 and $(m - M) = 29.98$,  that
corresponds to the distance $D = 9.9\pm 0.7$ Mpc.

\vspace{0.1cm}

\begin{center}
{\large \bf Metallicity distribution in the galaxy disk}
\end{center}
Since the location of the red supergiant branch at CMD depends on their
metallicity which can be well seen in the theoretical isochrones (Bertelli
et al., 1994), we used this effect for studying metallicity distribution
of stars along the radius of NGC~5194. For relative measurement of red
supergiants metallicity we use average colour indices relative to the colour
index of the blue supergiants branch.
Such approach makes the transition to absolute measurements of metallicity unnecessary.

 Red supergiants colour indices have been obtained in seven
ring-shaped regions (Fig.~6), the stars being selected in the band $I = 23.0
\pm0.25 (M_I =-7)$. Variation of the average colour value along the galaxy
radius is shown in Fig.~7. A sharp drop of metallicity in the central galaxy
region, slow variation in spiral arms and the second sharp drop in the
periphery of the thin disk can be well seen in this figure. The obtained results
are in agreement with the metallicity variations of the young stars of
our Galaxy (Shaver et al., 1983) and other galaxies (Marquez et al., 2002).

\vspace{0.1cm}

\begin{center}
{\large \bf Distribution of stellar number density along the galaxy radius}
\end{center}
Having picked out young supergiants and the stars of intermediate age (AGB)
on CM diagram (Fig. 3), we have built of stellar number density
along the radius of NGC~5194 (Fig. ~8). In spite of some fluctuation, the
distribution of stellar number density of supergiants corresponds to the empirical
model of the stellar structure of the spiral galaxies (Tikhonov et al., 2005),
i.e. a small gradient of the number density is observed in the thin disk
(up to Radious = of 8 kpc) and a sharp drop out of it. The AGB stars are
distributed in the disk more evenly and they stretch along the radius further
than supergiants, which also corresponds to the model of stellar structure of
galaxies (Galazutdinova, 2005). A disagreement with the model  is observed in
the central galaxy region, where AGB number density decreases although
according to the model it must increase. Probably, the AGB stars deficiency
in the central galaxy regions may be a result of interaction of galaxies.

\vspace{0.1cm}

\begin{center}
{\large \bf Stellar content of the tidal formations of NGC~5195}
\end{center}
In the interaction of galaxies tidal formations called "feathers" have been
formed around NGC~5195. A question about their nature had been  raised long ago
(Toomre \& Toomre, 1972), but only deep ACS/WFC images helped us to clarify
this problem. By choosing different age stars and drawing their spatial
distribution, we have found that young supergiants are seen in the tidal arm
and in star formation regions of NGC~5194. The red giants are evenly
distributed around  both galaxies and AGB stars are concentrated in
"feathers" (Fig.~9). Having chosen the region of maximum concentration of AGB
stars closed to NGC~5195, we have constructed CM diagram for this region and
imposed isochrones Bertelly et al. (1994) with $z=0.02$ on it. The age of
the youngest  AGB stars is equal to 80 million years, which agrees with
the model of repeated approaches of galaxies (Salo \& Laurikainen, 2000),
but  age distribution of AGB stars has maximum at 400--500 Myr, which
agrees with model of Toomre \& Toomre (1972).

\vspace{0.1 cm}

\begin{center}
{\large \bf References}
\end{center}

\bigskip
\noindent  Baron E.  et al., MNRAS, 279, 799, 1996

\noindent  Bertelli G. et al.,  A\&A, 106, 137, 1994

\noindent  Durrell P. R. et al., ApJ, 582, 170, 2003

\noindent  Feldmeier J.J., ApJ,  479, 231, 1997

\noindent Galazutdinova O.A., PhD Thesis, Nizhnij Arkhyz, 2005

\noindent Georgiev Ts. B.  et al., 1990, Astron. Letters, 16, 979, 1990

\noindent  Iwamoto K.  et al., ApJ, 437, 115, 1994

\noindent  Marquez I. et al., A\&A, 393, 389, 2002

\noindent  Salo H. \& Laurikainen E., MNRAS, 319, 377, 2000

\noindent  Sandage A. \& Tammann G.A., ApJ, 194, 559, 1974

\noindent  Shaver P.A. et al., MNRAS, 204, 53, 1983

\noindent  Takats K. \& Vinko J., MNRAS, 372, 1735, 2006

\noindent  Tikhonov N.A. et al., 2005, A\&A, 431, 127

\noindent  Toomre A. \& Toomre J., ApJ, 178, 623, 1972

\newpage
\begin{figure}[h]%
\centerline{\includegraphics[width=11cm, bb=13 15 157 136,clip]{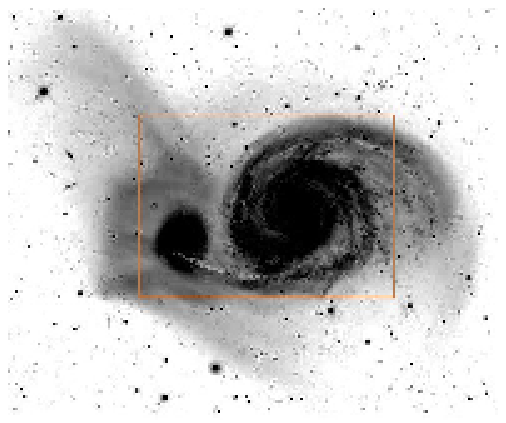}}%
\caption{ }
\end{figure}

\begin{figure}[h]%
\centerline{\includegraphics[width=11cm, bb=15 15  321 233,clip]{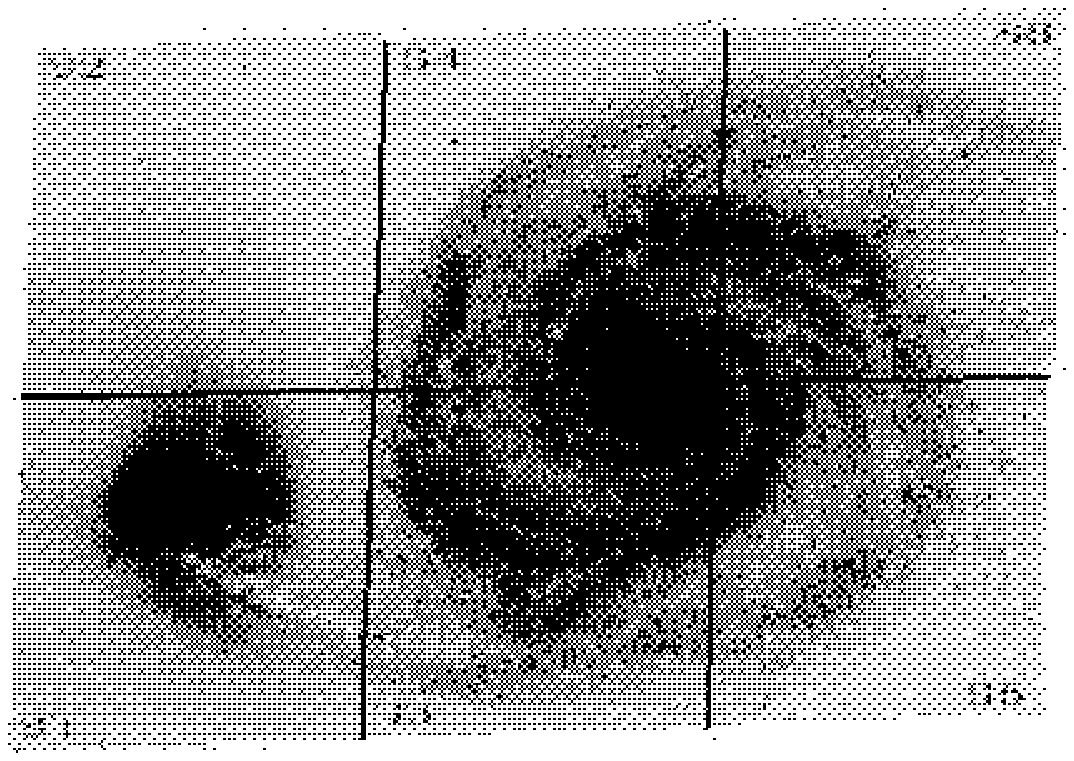}}%
\caption{}
\end{figure}

\begin{figure}[h]
\centerline{\includegraphics[width=11cm, bb=29 26 1160 1096,clip]{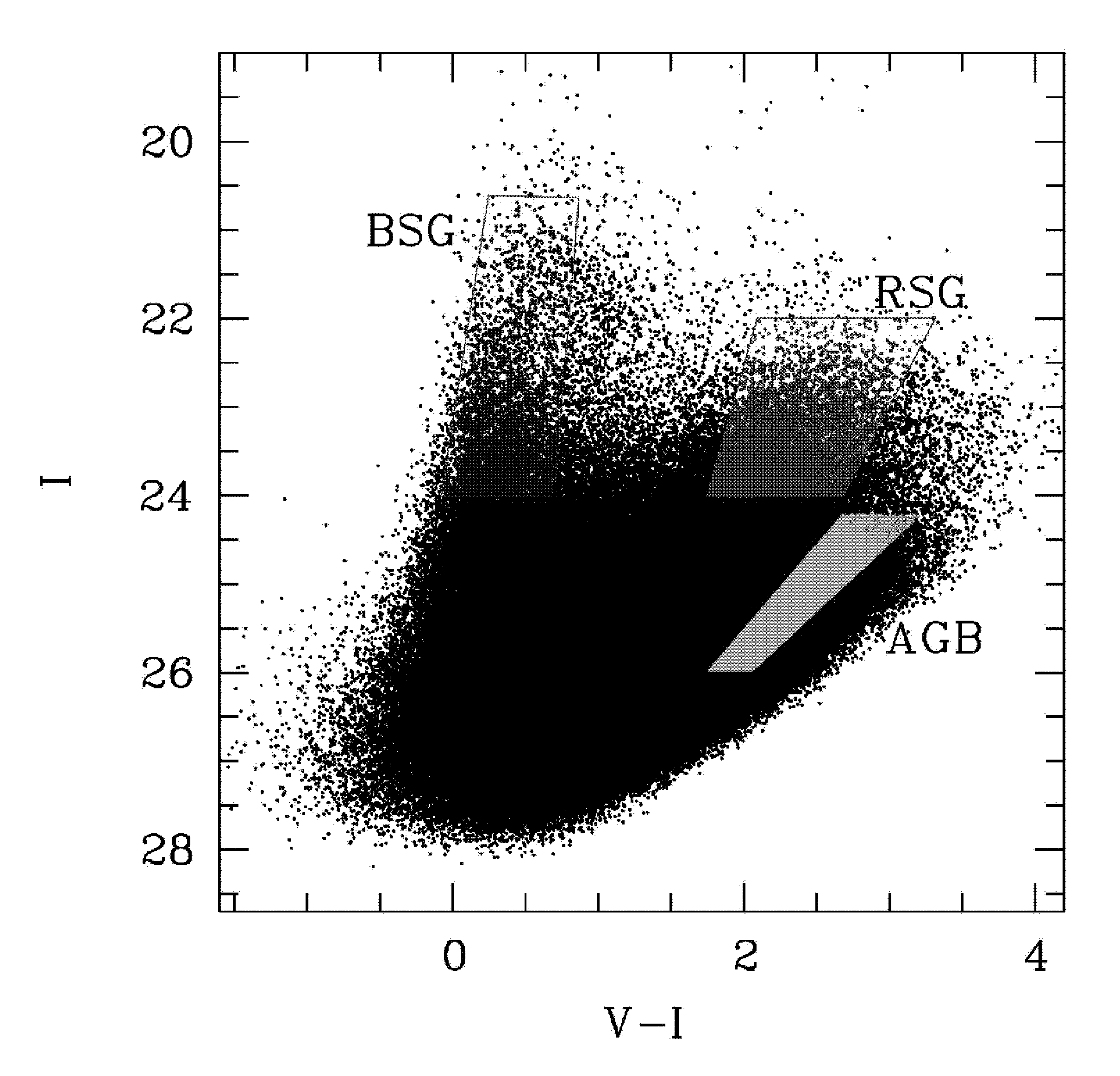}}%
\caption{}
\end{figure}

\begin{figure}[h]
\centerline{\includegraphics[width=11cm, bb=40 24 1145 1096,clip]{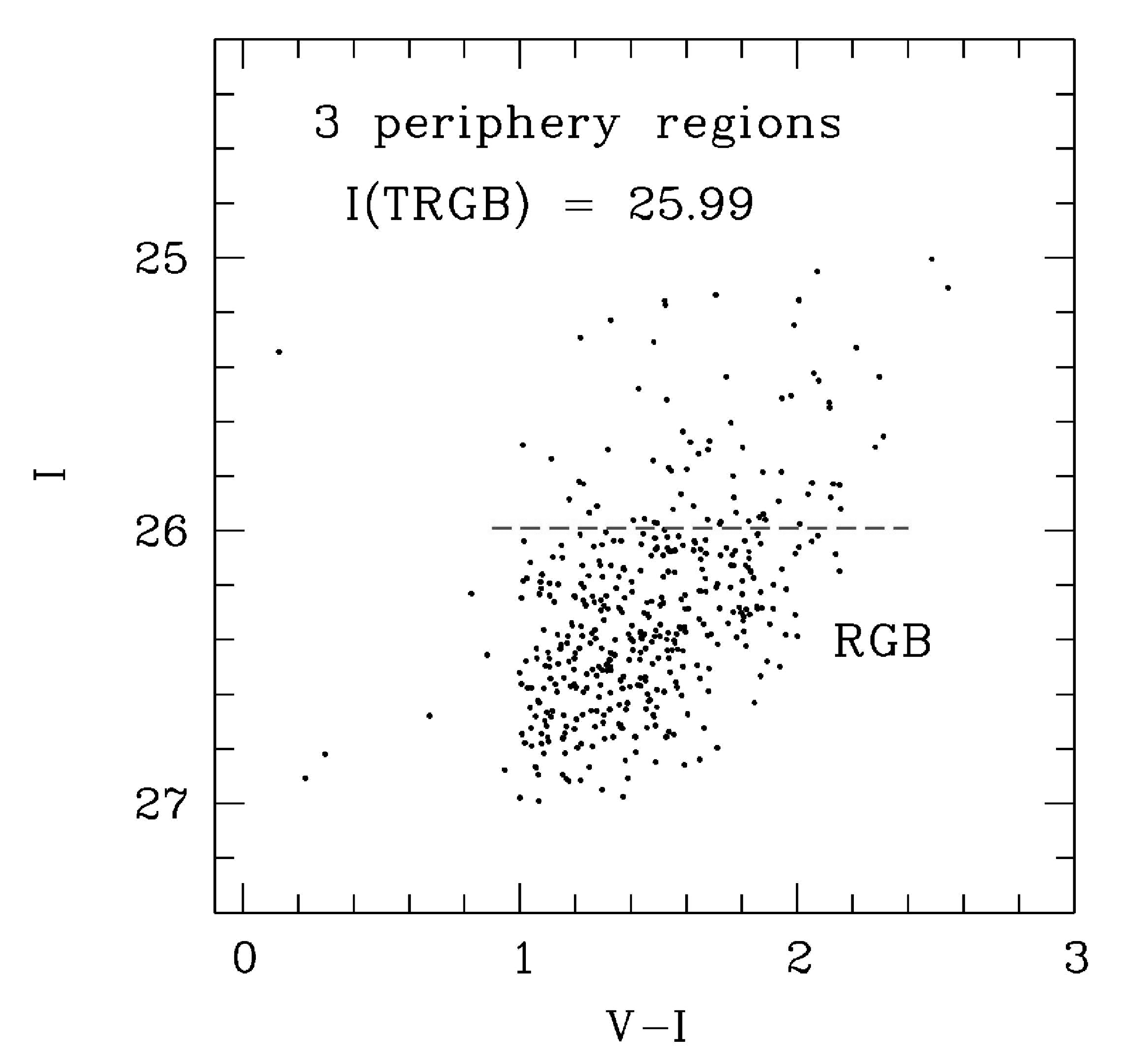}}%
\caption{}
\end{figure}

\begin{figure}[h]%
\centerline{\includegraphics[width=11cm, bb=53 31 720 656,clip]{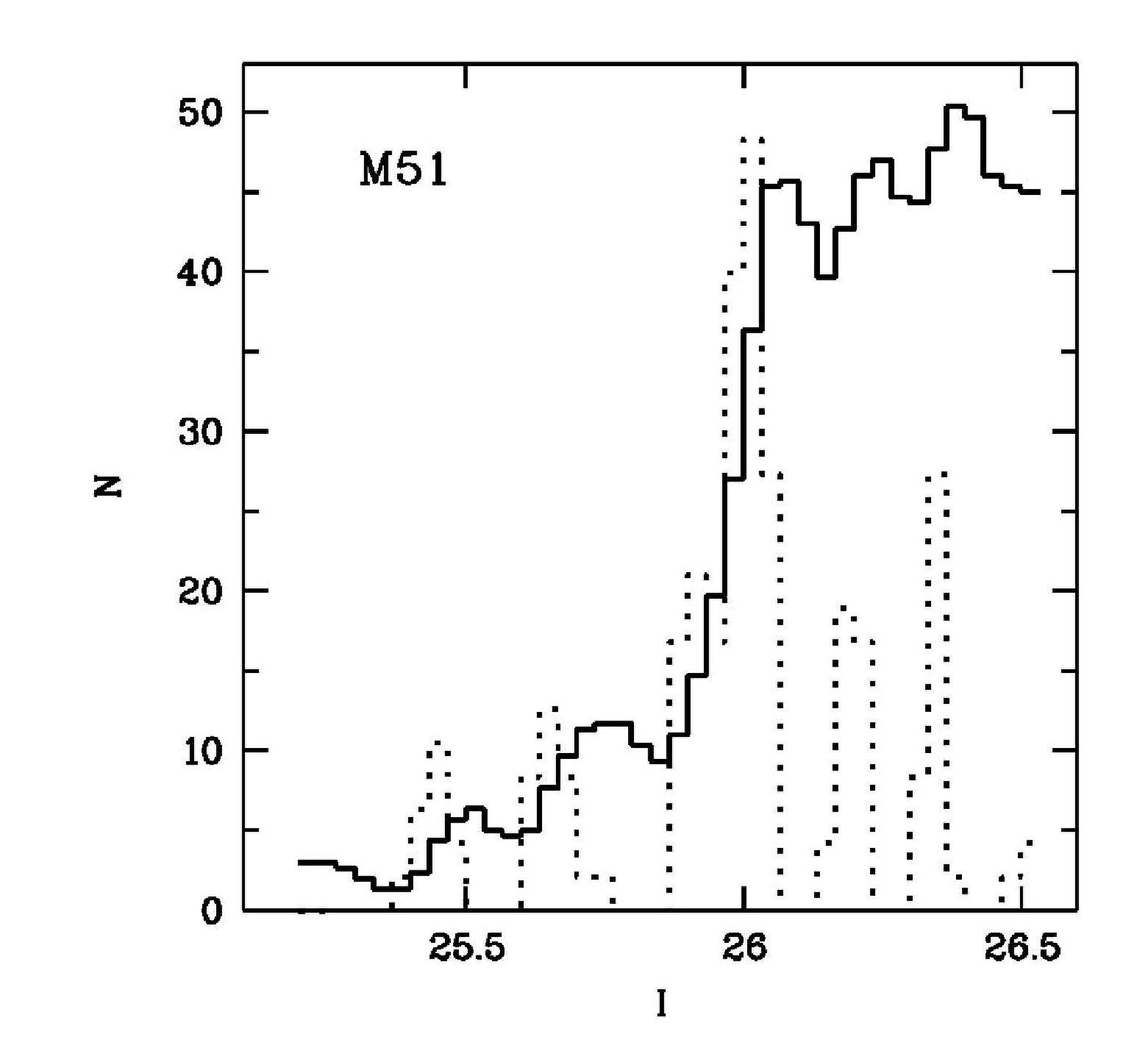}}%
\caption{}
\end{figure}

\begin{figure}[h]%
\centerline{\includegraphics[width=11cm, bb=18 22 320 237,clip]{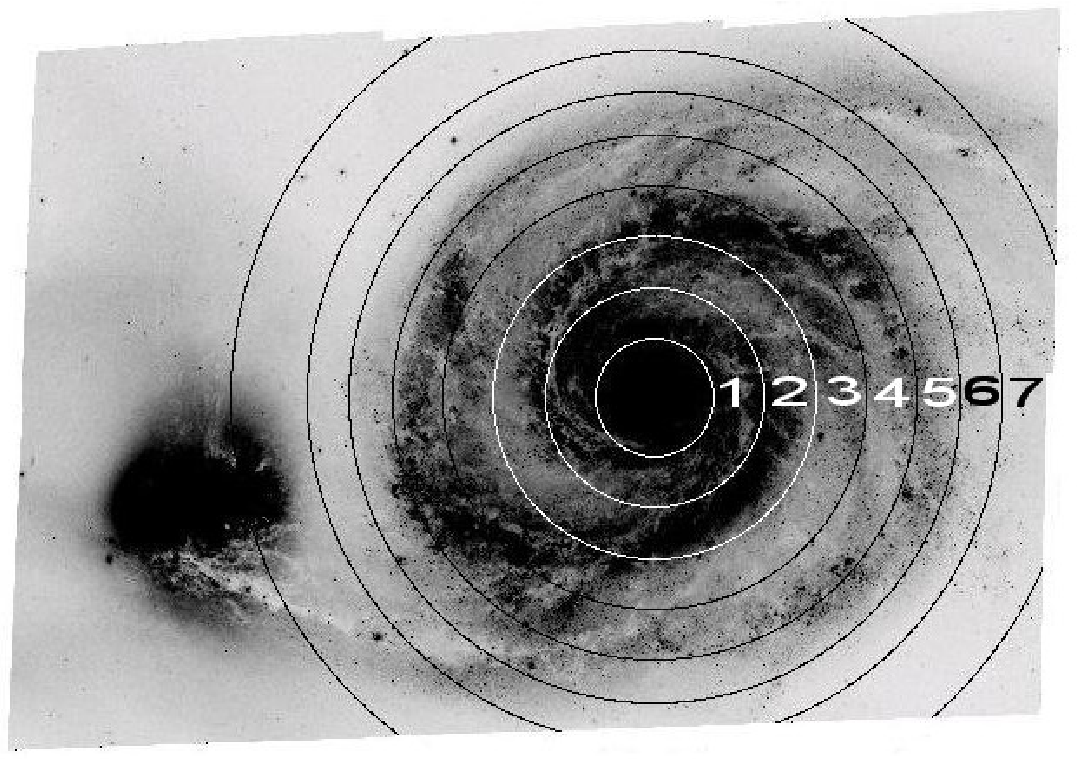}}%
\caption{ }
\end{figure}

\begin{figure}[h]%
\centerline{\includegraphics[width=11cm, bb=42 24 1199 843,clip]{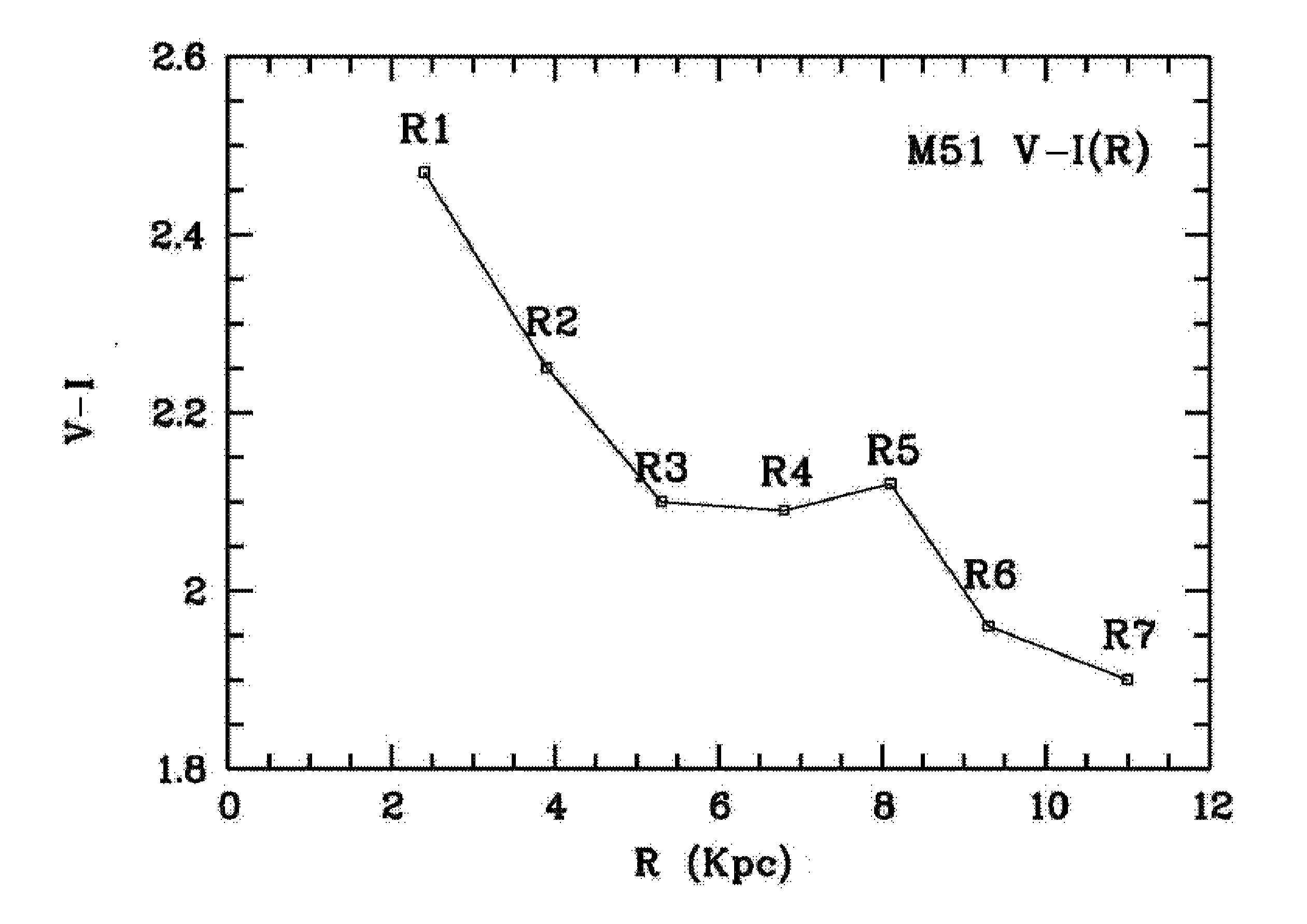}}%
\caption{ }
\end{figure}

\begin{figure}[h]%
\centerline{\includegraphics[width=11cm, bb=30 22 787 568,clip]{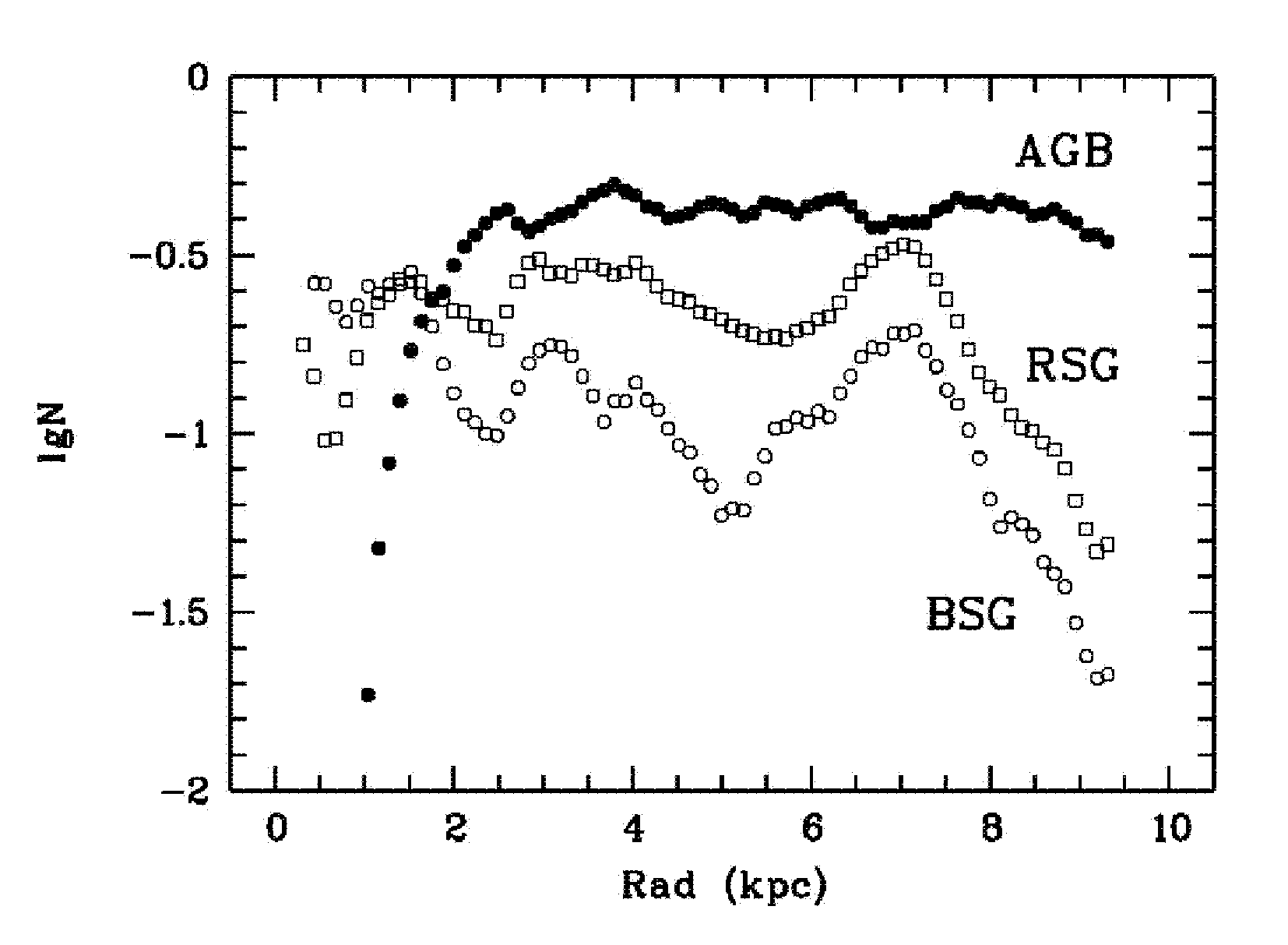}}%
\caption{}
\end{figure}

\begin{figure}[h]%
\centerline{\includegraphics[width=11cm, bb=24 22 632 869,clip]{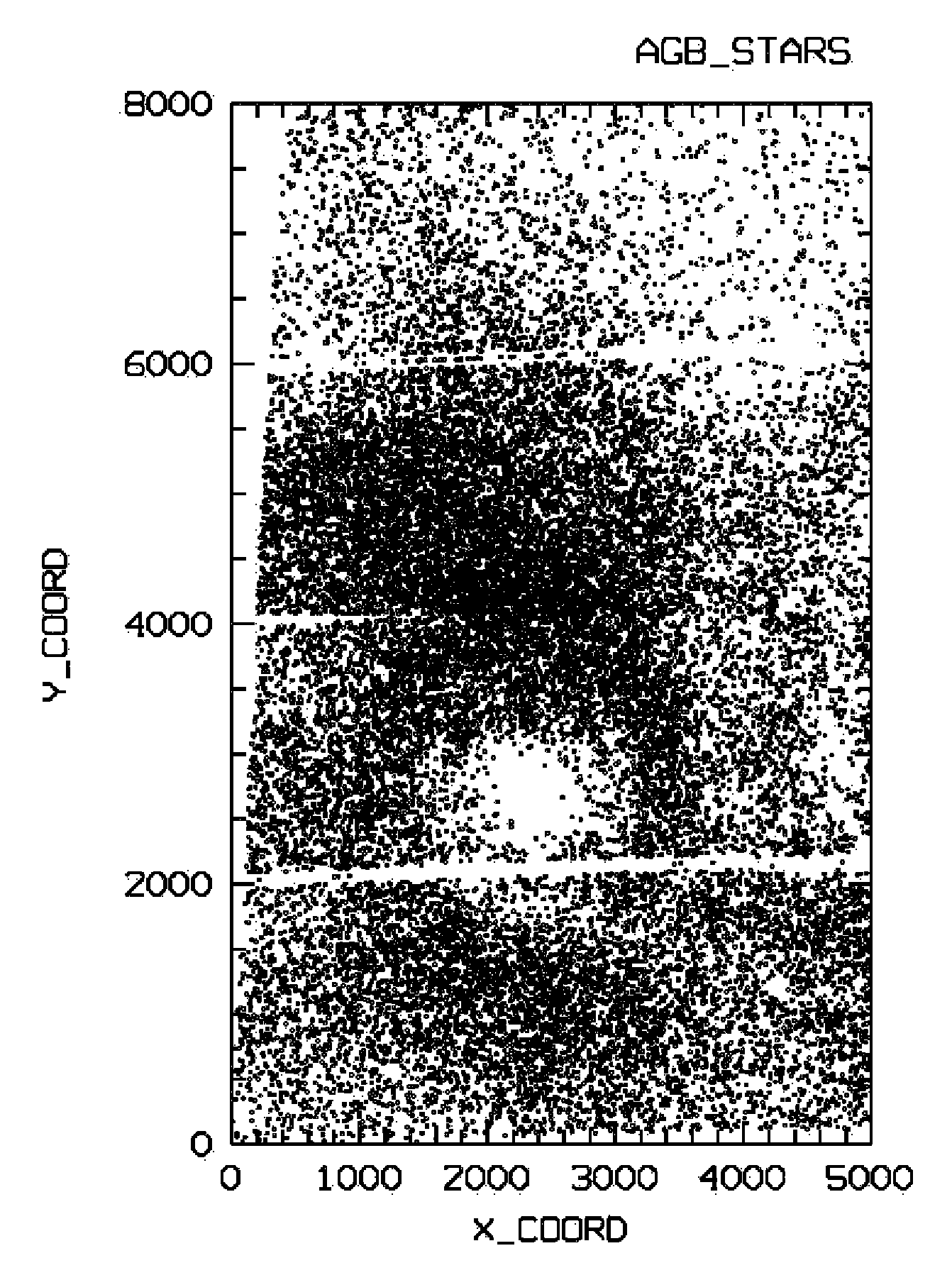}}%
\caption{}
\end{figure}

\end{document}